\documentclass[sigconf]{acmart}

\renewcommand\footnotetextcopyrightpermission[1]{} 
\usepackage{graphicx}
\usepackage{booktabs}
\usepackage{stfloats}
\usepackage[titletoc]{appendix}

\usepackage{algorithm}
\usepackage{algorithmic}
\usepackage{lipsum}
\usepackage{graphicx}
\usepackage{amsmath}
\usepackage{amssymb}

\AtBeginDocument{%
  }

\setcopyright{acmlicensed}
\copyrightyear{2024}
\acmYear{2024}

\acmConference[KDD '2026]{Make sure to enter the correct
 conference title from your rights confirmation emai}{TBD}{TBD}

\acmISBN{978-1-4503-XXXX-X/18/06}

\begin{document}
\title{VALUE: Value-Aware Large Language Model for Query Rewriting via Weighted Trie in Sponsored Search
}

\author{Xiao Zhang\textsuperscript{1}, Guanyu Chen\textsuperscript{1}, Boyang Zuo$^\ast$\textsuperscript{2}, Feng Li\textsuperscript{1}, Pengjie Wang\textsuperscript{1}, Jian Xu\textsuperscript{1}, Bo Zheng\textsuperscript{1}}
\email{{zx142853, cgy270526, adam.lf, pengjie.wpj, xiyu.xj, bozheng}@alibaba-inc.com, zuoby22@mails.tsinghua.edu.cn}
\affiliation{%
  \institution{\textsuperscript{1}Alibaba Group,  \textsuperscript{2}Tsinghua University}
  \city{Beijing}
  \country{China}
}

\thanks{$^\ast$This author contributed to this work during his internship at Alimama.}


\begin{abstract}
   Query-to-bidword (i.e., bidding keyword) rewriting is fundamental to sponsored search, transforming noisy user queries into semantically relevant and commercially valuable keywords. Recent advances in large language models (LLMs) improve semantic relevance through generative retrieval frameworks, but they rarely encode the commercial value of keywords. As a result, rewrites are often semantically correct yet economically suboptimal, and a reinforcement learning from human feedback (RLHF) stage is usually added after supervised fine-tuning (SFT) to mitigate this deficiency. However, conventional preference alignment frequently overemphasize the ordering of bidword values and is susceptible to overfitting, which degrades rewrite quality. In addition, bidword value changes rapidly, while existing generative methods do not respond to these fluctuations. To address this shortcoming, we introduce \textbf{VALUE} (\textbf{V}alue-\textbf{A}ware \textbf{L}arge language model for q\textbf{U}ery rewriting via w\textbf{E}ighted trie), a framework that integrates value awareness directly into generation and enhances value alignment during training. VALUE employs the Weighted Trie, a novel variant of the classical trie that stores real-time value signals for each token. During decoding, the framework adjusts the LLM’s token probabilities with these signals, constraining the search space and steering generation toward high-value rewrites. The alignment stage uses a fine-grained preference learning strategy that emphasizes stable, high-value differences and down-weights noisy or transient fluctuations, thereby improving robustness and reducing overfitting. Offline experiments show that VALUE significantly outperforms baselines in both semantic matching and value-centric metrics. Online A/B tests further revealed that our Revenue Per Mille (RPM) metric increased by 1.64\%. VALUE has been deployed on our advertising system since October 2024 and served the Double Eleven promotions, the biggest shopping carnival in China. 
\end{abstract}

\renewcommand{\shortauthors}{Xiao Zhang}


\maketitle

\section{Introduction}

Sponsored search is a crucial element of modern search engines \cite{fain2006sponsored}. Its effectiveness depends on accurately mapping user queries to relevant, high-value bidwords auctioned by advertisers. Query texts, however, are linguistically heterogeneous: they may contain imprecise phrasing, spelling errors, or synonyms. These variations obscure user intent and reduce matching accuracy. For instance, the query \emph{“drinks for fat person”} should be reformulated as \emph{“low-sugar healthy drinks”} before it can align with advertiser keywords.
Meanwhile, the rapid expansion of e-commerce has further diversified query patterns. Current search behavior includes long-tail expressions, ambiguous product descriptions, and multi-intent requests, each linked to distinct advertisement sets with substantially different commercial value.
Addressing these challenges requires query rewriting and keyword matching frameworks that incorporate both semantic equivalence and value awareness. Such systems must improve intent recognition while simultaneously optimizing for the expected value of candidate bidwords.



In recent years, generative query rewriting has garnered significant attention, with most state-of-the-art approaches following a three stage paradigm comprising supervised fine-tuning (SFT), value alignment, and Trie-constrained decoding \cite{taobaowww,chen2025iterqr,wang2024one,liu2025multi}.
In the SFT stage, domain-specific knowledge is injected into a large language model (LLM) through carefully curated rewriting corpora. The model then leverages its powerful semantic understanding, reasoning, and summarization capabilities to transform complex, noisy user queries into semantically canonical forms.
During value alignment, reinforcement learning methods such as Proximal Policy Optimization (PPO), Direct Preference Optimization (DPO), Knowledge Transfer Optimization (KTO), and ORPO \cite{ouyang2022training, rafailov2024direct, ethayarajh2024kto, hong2024orpo} adjust the model’s preferences so that the generated rewrites maximize not only semantic fidelity but also economic objectives including click-through rate and conversion likelihood.
In the final generation stage, Trie-based decoding limits the output space to valid keyword candidates, ensuring syntactic correctness and retrieval feasibility.
Historically, research has largely concentrated on better knowledge injection during SFT and on incorporating relevance, fluency, and factual consistency during alignment. Despite these advances, current systems remain inadequate in value awareness:

\begin{enumerate}
\item A fundamental misalignment exists between the commercial value ranking of keywords (driven by bid price and conversion potential) and their semantic relevance ranking in query rewriting. This divergence implies that semantically optimal rewrites rarely coincide with maximal-value keywords, creating inherent tension for generative rewriting systems. Recent frameworks like MoBGM \cite{liu2025multi} address this by jointly optimizing value and fidelity during alignment, yet require delicate balancing of training dynamics (e.g., reward weighting) to prevent objective dominance, which significantly increasing complexity.

\item Dynamic bid adjustments further cause real-time keyword value fluctuations. Consequently, value rankings observed during training become obsolete at inference, creating a temporal mismatch that undermines models relying on static/ historical signals. Current systems thus lack real-time value awareness, limiting adaptability to market dynamics and compromising economic optimality.
\end{enumerate}

\begin{figure}[t]
    \centering
    \includegraphics[width=0.5\textwidth]{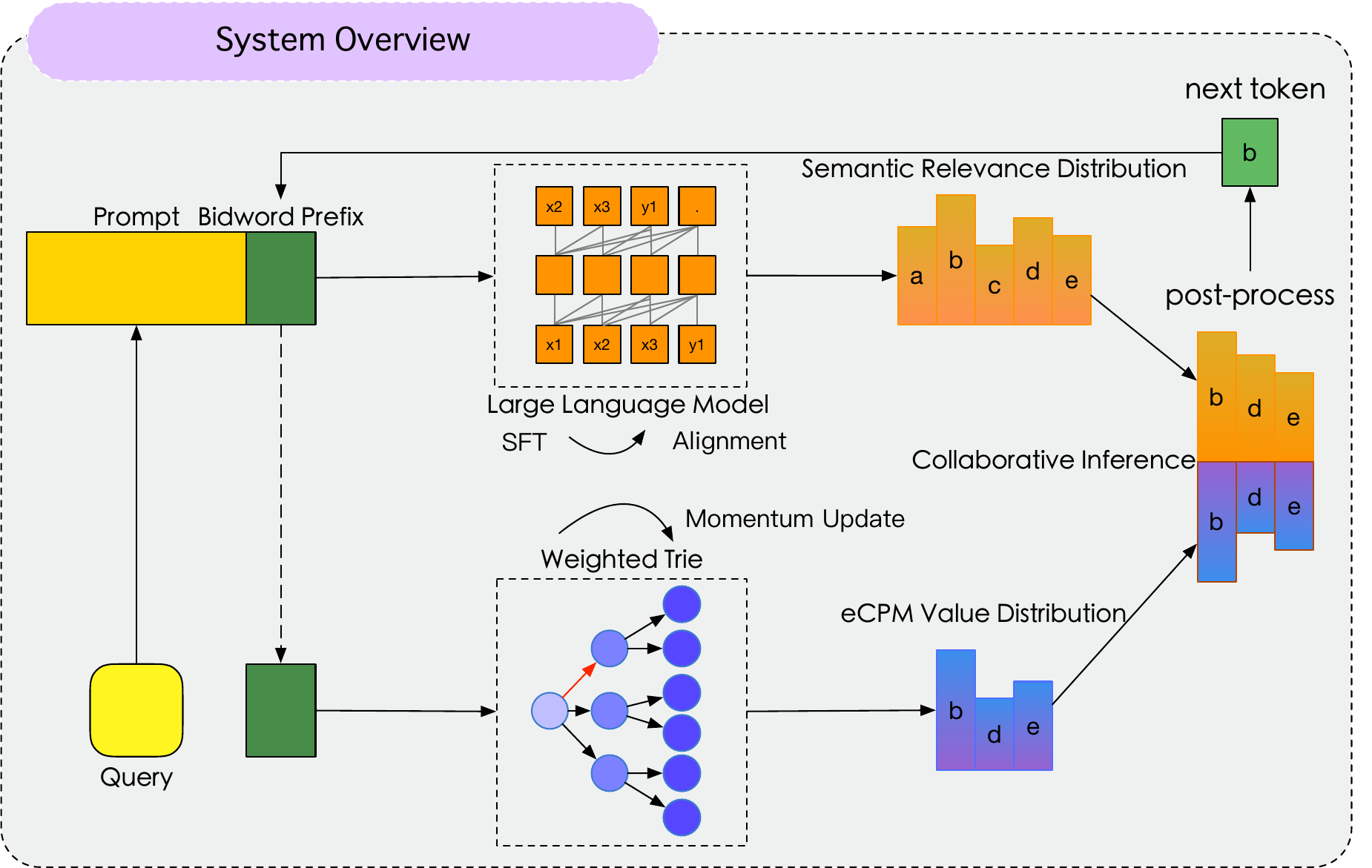}
    \caption{System Overview of our proposed VALUE Framework includes three parts: "SFT and Alignment", "Weighted Trie Construction and Weight Momentum Update", and "Collaborative Inference". During the LLM decoding process, we derive a value distribution from a Weighted Trie to constrain the generation space, and subsequently merge this with another distribution to produce the final output distribution.}
    \Description{}
    \label{fig:sysoverview}
\end{figure}

To address the aforementioned challenges, we propose shifting the focus of value integration from the training phase to the generation phase, where bidword value can be dynamically incorporated without sacrificing semantic relevance. We introduce a novel \textit{reward-guided inference} framework, named as \textbf{VALUE}, and design a lightweight \textit{Weighted Trie} structure that embeds bidword value directly into the decoding process, as illustrated in Figure~\ref{fig:sysoverview}.
Specifically, during Trie construction, we annotate each node with downstream bidword value information, providing token-level commercial signals along the generation path. This allows the model to reason over both linguistic plausibility and economic return at each decoding step. To account for real-time bid fluctuations, we further develop a \textit{dynamic Trie update strategy} that refreshes value annotations on the fly, ensuring the system remains responsive to changes in keyword valuation during inference.
At each decoding step, the model queries the Weighted Trie for value-based rewards and modulates the token probability distribution accordingly. By maximizing cumulative output value during inference, this process decouples economic optimization from semantic learning, mitigating catastrophic forgetting and overfitting associated with value-centric training objectives. The framework therefore supports an explicit trade off between rewriting relevance and commercial utility.

To further bias generation toward consistently high-value bidwords, we refine DPO with adaptive loss weighting. The revised loss amplifies preference pairs exhibiting large, stable value gaps while down-weighting pairs subject to transient noise. This selectivity reduces sensitivity to unstable signals and improves generalization to long-tail and dynamically priced keywords.
Overall, VALUE achieves a balanced integration of semantic fidelity and economic efficiency, bridging the gap between intent understanding and revenue maximization in generative query rewriting.

To evaluate the effectiveness of the proposed method, we conduct comprehensive offline experiments on historical online logs. Compared with the currently deployed production system, our approach achieves a 9.3\% improvement in semantic relevance. Furthermore, large-scale online A/B testing demonstrates consistent gains in key business metrics: consumption increases by 1.53\%, and revenue per thousand impressions (RPM) rises by 1.64\%. Given these significant and sustained improvements, our method has been fully integrated into the live serving pipeline.
Our contributions can be summarized as follows:

\begin{enumerate}
\item \textbf{A pioneering reward-guided collaborative inference framework for query rewriting:} We propose a novel generative rewriting framework that synergistically combines large language models (LLMs) with a value-augmented Weighted Trie, enabling dynamic, lightweight value awareness while preserving semantic fidelity, achieving a principled balance between relevance and commercial utility.
\item \textbf{Innovative Trie structure with embedded value propagation:} We redesign the classical Trie data structure by annotating each node with downstream bidword value information, providing token level commercial-value signals during generation. To our knowledge, this is the first work to transform the traditionally Trie-based constraint decoding into an active value-awareness mechanism within generative rewriting, allowing real-time adaptation to fluctuating bidword value through a dynamic update strategy.
\item \textbf{Fine-grained reward optimization in post-training alignment:} We introduce an enhanced preference learning algorithm that selectively emphasizes bidword pairs with large and stable value gaps, while down-weighting those susceptible to transient bid noise. This selective focus reduces noise exposure and mitigates overfitting risks, leading to more robust and generalizable value modeling.
\end{enumerate}

\section{Related Works}

\subsection{Query Rewriting}
The process of query rewriting, often termed query expansion or reformulation, is essential in enhancing e-commerce search technologies. This approach can be generally classified into two classes: discriminative methods and generative methods.

\textbf{Discriminative methods} frame query rewriting as a retrieval task. Pseudo-relevance feedback techniques \cite{xu,t6,t31,t40} expand queries by extracting terms from top-ranked initial documents, blending global corpus statistics with local feedback. While effective for lexical mismatch, these methods remain vulnerable to semantic drift from noisy top results. To mitigate this, curated thesauri have been proposed as candidate sources \cite{t5,t22}, though their efficacy critically depends on thesaurus quality.
Alternate approaches \cite{t3,t8,t18} leverage search logs to generate rewrites, using historically similar terms as expansions. However, frequency bias in click data skews training toward head queries, yielding over-optimized embeddings that neglect long-tail queries. Dense retrieval models further struggle with distribution shifts, requiring costly mitigation strategies (e.g., meta-learning \cite{taobaowww}) for long-tail generalization. Consequently, such methods inherently underperform for infrequent queries.

\textbf{Generative methods}, on the other hand, focus on the direct transformation of queries into bidwords \cite{baiduwww, lee2018rare, qi2020prophetnet, mohankumar2023unified, wang2024one, liu2025multi, chen2025iterqr}. Most of them adopt a standard pipeline of supervised fine-tuning, value alignment, and Trie-based decoding, which constrains outputs to bidwords that are both semantically relevant and commercially valuable. Earlier researches have improved supervised fine-tuning through prompt engineering to better exploit LLM reasoning and summarization, and has extended value alignment to incorporate objectives such as relevance and factual consistency. Nonetheless, existing systems still struggle to balance semantic fidelity with economic value and fail to adapt to real-time bidword fluctuations. Our approach, as detailed in the following sections, effectively tackles this challenge by introducing innovative techniques that not only ensure semantic relevance but also optimize for value.

\subsection{Preference Alignment}
Reinforcement Learning from Human Feedback (RLHF) aligns LLMs with human preferences by applying the Bradley–Terry model \cite{bradley1952rank} to estimate pairwise preferences \cite{christiano2017deep, bai2022training, ouyang2022training}. A typical RLHF pipeline comprises three stages: supervised fine tuning \cite{zhou2024lima, taori2023stanford, conover2023free, ding2023enhancing}, reward model training \cite{gao2023scaling, chen2024odin, havrilla2024glore, leike2018scalable}, and policy optimization \cite{schulman2017proximal, anthony2017thinking}. Proximal Policy Optimization (PPO) \cite{schulman2017proximal} is widely used in the final stage. RLHF has been applied to reduce toxicity \cite{amini2024direct, korbak2023pretraining}, improve safety \cite{dai2023safe}, enhance helpfulness \cite{wei2024long}, and strengthen reasoning \cite{havrilla2024teaching}. However, challenges arise across the pipeline, from preference data collection to model training \cite{casper2023open}, and RLHF can introduce biases such as excessively long outputs \cite{dubois2024length, singhal2023long, wang2023far}.

Alternative alignment methods avoid explicit reward model training. Direct Preference Optimization (DPO) \cite{rafailov2024direct}, Kahneman Tversky Optimization (KTO) \cite{ethayarajh2024kto}, Simple Preference Optimization (SimPO) \cite{meng2024simpo}, and Monolithic Preference Optimization (ORPO) \cite{hong2024orpo} optimize directly on preference pairs. SimPO and ORPO further remove the need for a reference model, simplifying training. Although these methods cannot sample preference pairs from an optimal policy, researchers augment data with outputs from a supervised fine-tuned model and employ iterative frameworks that periodically update the reference. Alignment as Reward Guided Search (ARGS) \cite{khanov2024args} integrates alignment into decoding instead of training. Despite their advantages, these alignment techniques remain computationally intensive and must be retrained whenever value criteria change.


RLHF methods are commonly employed in generative query rewriting to infuse bidwords value information. However, a weak correlation between bidwords value ranking and relevance ranking often arises during rewriting. This misalignment leads to overfitting and degraded rewriting relevance in value-aware generative models. Our proposed VALUE framework addresses these issues through a dual-pronged approach: (1) leveraging the Weighted Trie for lightweight, real-time value awareness during generation, and (2) mitigating catastrophic forgetting of SFT-injected knowledge via optimized value preference alignment.

\section{Methodology}

\begin{figure*}[t]
    \centering
    \includegraphics[width=\textwidth]{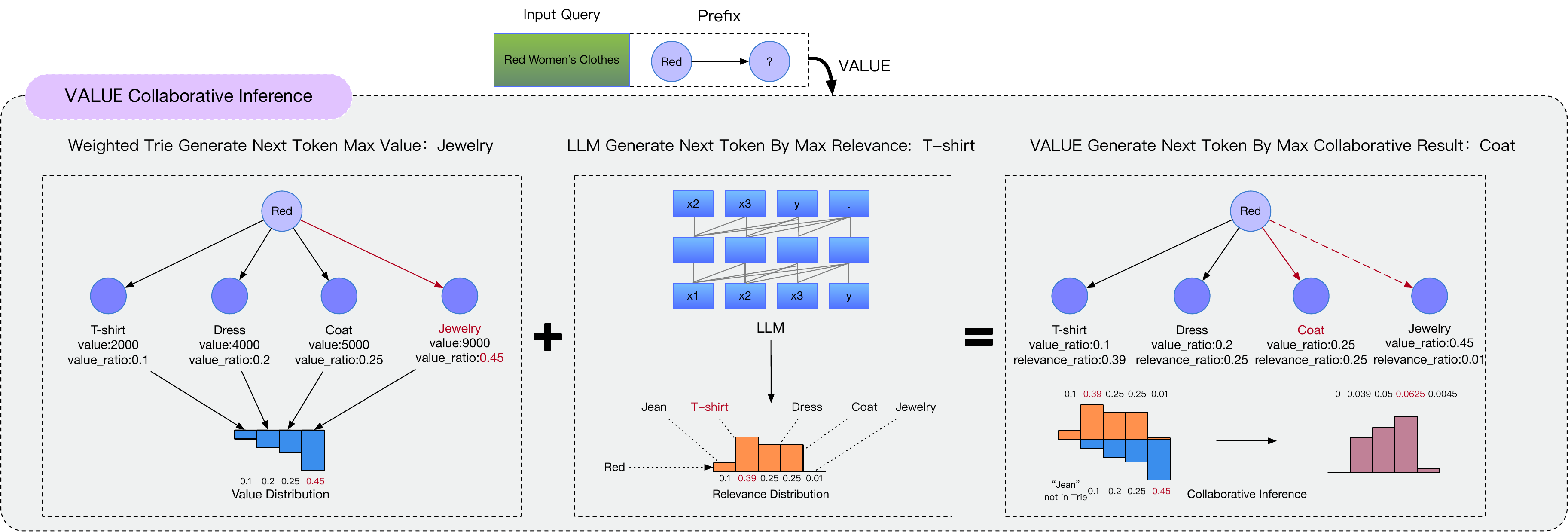}
    \caption{Collaborative Inference of VALUE framework. When generating the next token, we obtain two output distribution from the LLM and Weighted Trie. We exclude tokens not present in the value distribution and adjust the probabilities of the remaining tokens. Then, we sample the next token from the modified distribution.} 
    \label{fig:col}
\end{figure*}

As illustrated in Figure~\ref{fig:sysoverview}, VALUE operates in two primary phases: (1) a training phase integrating Supervised Fine-Tuning (SFT) with value preference alignment, and (2) a collaborative inference phase utilizing the Weighted Trie structure. During the SFT phase, preprocessed and relevant query-bidword pairs are employed to inject domain-specific rewriting knowledge for sponsored search into LLMs. Subsequently, we present our lightweight value-aware inference framework and the optimized value preference alignment method in detail.

We first introduce the construction of Weighted Trie and weight momentum update. Then, we will introduce collaborative inference based on Weighted Trie and the  post-training alignment method we proposed in training phase. Finally, we will describe how VALUE is deployed for online serving.

\subsection{Weighted Trie}

\subsubsection{Definition and Construction of Weighted Trie}
Weighted Trie is an renovation of the traditional trie data structure. Each node in the Weighted Trie, referred to as a \textbf{WeightedTrieNode}, contains additional attributes to store weight-related information. The primary attributes of a WeightedTrieNode are:

\begin{itemize}
    \item \textbf{children}: A dictionary where the keys are integers representing tokens in LLM's vocabulary, and the values are child nodes of type \textit{WeightedTrieNode}.
    \item \textbf{mean\_v}: A floating-point number representing the average eCPM value of the node's children, where eCPM (effective Cost Per Mille) is the bidword value calculated offline.
    \item \textbf{max\_v}: A floating-point number representing the maximum eCPM value among the node's children.
    \item \textbf{is\_word}: A boolean value indicating whether the node represents a complete word.
\end{itemize}

The construction of the Weighted Trie involves initializing the root node, inserting bidwords with their corresponding values, and updating the weights of the nodes. The pseudocode in Algorithm 1 demonstrates the construction process.

\begin{algorithm}
\caption{Construct Weighted Trie}
\begin{algorithmic}
\STATE \textbf{Input:} List of pairs $(\textit{bidword}, \textit{ecpm})$
\STATE \textbf{Output:} Weighted Trie
\STATE Initialize weighted trie: $\mathit{weighted\_trie}$
\FOR{each $(\textit{bidword}, \textit{ecpm})$ \textbf{in} list}
    \STATE $\mathit{token\_ids} \leftarrow \text{tokenizer}(\textit{bidword})$
    \STATE Insert $\mathit{token\_ids}$ into the trie:
    \FOR{each $\mathit{token\_id}$ \textbf{in} $\mathit{token\_ids}$}
        \STATE Traverse or create node for $\mathit{token\_id}$
    \ENDFOR
    \STATE Set $\mathit{leaf\_node.mean\_v} \leftarrow \textit{ecpm}$
    \STATE Set $\mathit{leaf\_node.max\_v} \leftarrow \textit{ecpm}$
\ENDFOR
\STATE Update weights in the Weighted Trie using post-order traversal:
\FOR{each \textit{node} \textbf{in} post-order}
    \STATE $\mathit{node.mean\_v} \leftarrow \frac{1}{N} \sum_{\mathit{child} \in \mathit{children}} \mathit{child.mean\_v}$ \\
    \STATE $\mathit{node.max\_v} \leftarrow \max_{\mathit{child} \in \mathit{children}} \mathit{child.max\_v}$
\ENDFOR
\end{algorithmic}
\end{algorithm}

\subsubsection{Weighted Trie Momentum Update}
Momentum updates are essential due to the frequent changes in bidding prices and the evolving bidword space. The process is delineated as follows:

\paragraph{Tokenization and Node Insertion:} Each bidword undergoes tokenization, converting it into a sequence of token IDs. These IDs facilitate the insertion or updating of nodes within the trie. The update process for a node is governed by:

\begin{equation}
Max\_V_{\text{new}} = \max(eCPM_{\text{new}}, Max\_V_{\text{old}})
\end{equation}

\begin{equation}
Mean\_V_{\text{new}} = \alpha \cdot eCPM_{\text{new}} \ + \  (1-\alpha) \cdot Mean\_V_{\text{old}}
\end{equation}

where $\alpha$ is the hyperparameter that control the update rate. If the node is absent, it is initialized as:

\begin{equation}
Max\_V_{\text{new}} = eCPM_{\text{new}},\  Mean\_V_{\text{new}} = eCPM_{\text{new}}
\end{equation}

\paragraph{Bottom-up Update:} Following node insertion, a bottom-up iteration is executed to update all intermediate nodes, ensuring that the entire trie structure accurately reflects the most recent eCPM values.

This method enables the trie to dynamically adjust to variations in the bidword space and eCPM values over time, preserving the framework's efficacy without necessitating frequent retraining, maintaining stability and performance in a dynamic e-commerce environment.

\subsection{Collaborative Inference}

As shown in Figure~\ref{fig:col}, when generating the \( t \)-th token, we insert the query into the prompt template and concatenate it with the bidword prefix(i.e. the first t-1 tokens) to form the input of LLM to get the output probabilities. We use generated bidword prefix to retrieve all child nodes from the trie. The value of these child nodes is computed using the following formula:
\begin{equation}
\mathcal{V}_k = \beta \cdot Max\_V_k + (1 - \beta) \cdot Mean\_V_k 
\end{equation}
where \(\beta\) is the tunable hyperparameter set to 0.5, allowing us to prioritize paths with higher average values or to explore bidwords with the maximum potential value, depending on our specific needs.

We then apply the softmax function to these values to obtain the normalized value \(\hat{\mathcal{V}}_k\). 
Next, we mask out all probabilities in the LLM-generated distribution that are not children of the previous generated token in the Weighted Trie. For the remaining probabilities, we adjust them by weighting with the normalized value \(\hat{\mathcal{V}}_k\) and a depth-dependent factor \(\theta\). The adjusted probability for each token \( k \) is given by:

\begin{equation}
p_{WT}(k \mid x_{<t}) = p_{{LLM}}(k \mid x_{<t}) \cdot (1 + \hat{\mathcal{V}}_k \cdot \theta_t)
\end{equation}
The fusion value \(p_{WT}(k \mid x_{<t})\) is an affine transformation of \(p_{{LLM}}(k \mid x_{<t})\) and \(\hat{\mathcal{V}}_k\), selected through a series of tests and comparisons with additive fusion, multiplicative fusion, etc.
where \(\theta_t \in \Theta\) is designed to vary with the depth \( t \) of the trie, playing a crucial role in balancing relevance and value in our model's output.  Specifically, \(\Theta\) assigns lower value weight to initial generated tokens and higher weight to later generated tokens. This design stems from two key observations: (1) Initial tokens primarily establish semantic relevance. Excessive value focus at this stage risks compromising query intent alignment. (2) Subsequent tokens refine bidword specificity. Once core semantics are anchored, increased value weighting allows precise optimization within the relevant search space. Thus, \(\Theta\) progressively shifts focus from relevance grounding to value maximization across the generation trajectory. Consequently, we derive a modified probability distribution \(p_{WT}(k \mid x_{<t})\), from which the next token is sampled as usual.

\subsection{LLM Post-training Alignment with Weighted Trie}

Although the Weighted Trie offers a lightweight mechanism for balancing relevance and commercial value of rewriting results, we seek to further enhance the LLM's generation probability for high-value bidwords. This is critical because, in scenarios where a query has numerous relevant bidwords, high-value candidates meeting the relevance threshold might be excluded from the rewrite set due to low generation probabilities, thereby limiting the effectiveness of the Weighted Trie. However, prevailing value alignment techniques, exemplified by DPO, often excessively prioritize bidwords value ranking. This over-emphasis can lead to model overfitting or catastrophic forgetting of the semantic knowledge acquired during SFT.

To address these challenges, we propose Weighted Direct Preference Optimization (WDPO). WDPO directly incorporates bidwords value into DPO, steering the model's attention toward bidword pairs exhibiting substantial value disparities. This approach effectively mitigates the negative impact of the value alignment phase on rewriting relevance.

The alignment process proceeds as follows:
\begin{enumerate}
\item Identify relevant bidwords from online logs and calculate their corresponding eCPM values for each query in our dataset.
\item Randomly sample two bidwords $(y_w, y_l)$ with differing eCPM values, ensuring:
\begin{equation}
|eCPM(y_w) - eCPM(y_l)| > \tau
\end{equation}
where $\tau$ is a predefined threshold, $y_w$ represents the bidword with higher eCPM, and $y_l$ the lower.
\item Process the query and selected bidwords through both policy and reference models to obtain probabilities:
$\pi_\theta(y_w|x)$, $\pi_\theta(y_l|x)$, $\pi_{ref}(y_w|x)$, and $\pi_{ref}(y_l|x)$.
\item Apply the modified DPO loss function:
\begin{equation}
    \begin{aligned}
        \mathcal{L}_{WDPO}(\pi_\theta) = -\mathbb{E}_{(x,y_w,y_l)\sim \mathcal{D}} \bigg[ & \log \sigma\bigg(\beta \log\frac{\pi_\theta(y_w|x)}{\pi_{ref}(y_w|x)} \\
        & - \beta \log\frac{\pi_\theta(y_l|x)}{\pi_{ref}(y_l|x)}\bigg) \cdot w\bigg]
    \end{aligned}
\end{equation}
where $w$ is a weight function based on KL divergence:
\begin{equation}
w = \exp(-KL([\pi_\theta(y_w|x), \pi_\theta(y_l|x)] \parallel [P(y_w), P(y_l)]))
\end{equation}
and $P(y_i)$ represents normalized eCPM values:
\begin{equation}
P(y_i) = \frac{eCPM(y_i)}{\sum_{j \in \{w,l\}} eCPM(y_j)}
\end{equation}
\end{enumerate}

\subsection{Online Depolyment}

We constructed a WeightedTrie using a dataset of 30 million bidwords. In real-world business scenarios, the growth of the bidwords is typically gradual and has negligible impact on query processing performance. As an alternative, a pruning strategy based on key metrics such as eCPM or historical number of clicks can be employed to maintain the bidword count within predefined performance bounds.

In the context of deploying LLMs for online serving, the balance between computational efficiency and latency is paramount. Our approach leverages the Qwen2-7B\cite{team2024qwen2} model for offline inference, targeting head queries that constitute 85\% of the page views (PV) in our system. This offline process generates the top 500 bidwords per query, which are subsequently cached for rapid access. For mid-tail and long-tail queries that do not benefit from the cached results, we employ the Qwen2-1.5B model in an online serving capacity. This model is optimized to deliver a latency of 50 milliseconds, efficiently producing the top 50 bidwords per query. In the future, we plan to integrate user behavior sequence data into our online serving framework to enhance personalized recommendation systems and gradually remove the offline part.

\section{Experiments}

\subsection{Datasets}

\noindent \textbf{Training Dataset:} For multi-task fine-tuning, we extracted 150 billion records from 30 days of online logs, followed by several rounds of data cleaning. Initially, we employed manual observation and regular expressions to filter out the majority of noisy queries. In the subsequent screening phase, we calculated the click-through rate (CTR) based on page views (PV) and clicks, retaining only those $\langle\texttt{query},\texttt{bidword}\rangle$ pairs associated with purchase behavior that meet the click-through rate requirements. We then applied a relevance model for further filtering. This relevance model is a fine-tuned version of BAAI/bge-large-zh-v1.5\cite{xiao2024c}, optimized for e-commerce scenarios. We truncated the number of bidwords for each query, retaining a maximum of 50 bidwords per query. After third round of filtering, we were left with 110 million records.

\noindent \textbf{Test Dataset:} Our test dataset was constructed by randomly sampling 30,000 queries from online logs. These queries were stratified into three categories: 40\% head queries, 40\% torso queries, and 20\% tail queries.

\subsection{Evaluation Metrics}

\noindent \textbf{Offline Metrics:} 
To comprehensively evaluate our query rewrite approach, we utilize several offline metrics:

\begin{enumerate}

\item \textbf{Hit Rate:} A precision-oriented metric measuring the alignment between generated rewrites and user-clicked bidwords. For a query set $\mathcal{Q}$, it is defined as:
\begin{equation}
    \text{hitrate@}K = \frac{\sum_{q \in \mathcal{Q}} | \mathcal{R}_q^K \cap \mathcal{C}_q |}{\sum_{q \in \mathcal{Q}} | \mathcal{C}_q |}
\end{equation}
where $\mathcal{R}_q^K$ represents the top-$K$ rewrite candidates for query $q$, and $\mathcal{C}_q$ denotes the set of clicked bidwords for $q$. The numerator counts total hits across all queries, while the denominator normalizes by the total clickable bidwords.

\item \textbf{Relevance Score:} A semantic alignment measure between rewritten queries and the original query intent, typically quantified via pretrained language model similarity (e.g., BERTScore).

\item \textbf{Spearman Rank Correlation ($\rho$):} A non-parametric measure of rank consistency between model-generated and ground-truth bidword rankings. For $n$ observations:
\begin{equation}
    \rho = 1 - \frac{6 \sum_{i=1}^n d_i^2}{n(n^2 - 1)}
\end{equation}
where $d_i$ is the rank difference for the $i$-th bidword pair. The coefficient $\rho \in [-1, 1]$ reflects the monotonicity of the predicted rankings.

\end{enumerate}

\noindent \textbf{Online Metrics:} To evaluate the model's performance, we use three key metrics: \textbf{cost}, which is the total advertiser expenditure for clicks within a specific traffic segment, \textbf{revenue per mille (RPM)}, which measures the revenue generated per thousand ad impressions and \textbf{Page View Relevance (PV rele)},

\subsection{Implementation Details}

\noindent \textbf{Supervised Fine-tuning:} We fine-tuned the Qwen2-7B model using a learning rate of 1e-5 and a cosine learning rate scheduler. The process was optimized with AdamW ($\beta_1 = 0.9$, $\beta_2 = 0.999$) and a weight decay of 0.001. Gradient accumulation was set to 4. We used a batch size of 128 and applied the DeepSpeed ZeRO Stage 2 parallelism strategy with bfloat16 precision. The training was performed on 32 NVIDIA H20 GPUs for a total of 12 hours. 

\noindent \textbf{Post-training Alignment:} For the post-training alignment phase, we use our fine-tuned Qwen2-7B model. The alignment phrase was carried out with a learning rate of 1e-6 and a batch size of 64. We set the $\beta$ parameter to 0.1. The training was executed on 16 NVIDIA H20 GPUs for a duration of 3 hours. 

\noindent \textbf{Construction and Update of Weighted Trie:} We built a Weighted Trie containing 30 million bidwords with an average length of 7.9 characters, which occupies approximately 2.7 GB of memory after serialized. Single-threaded construction finishes in 56 minutes on a 32-core CPU machine, but multi-threaded optimization reduces this time to roughly 20 minutes. We used a message queue for streaming updates to the Weighted Trie, achieved minute-level update latency.

\subsection{Offline Experiments}

\begin{table*}[t]
\centering
\caption{Comparative Performance Analysis of Various Models. The eCPM metric is used as an approximate estimation of reward, serving as a reference for the effectiveness of Weighted Trie collaborative inference.}
\label{table:performance}
\begin{tabular}{clcccccc}
\toprule
Experiment Number & Model & hitrate@50 & hitrate@500 & Spearman's $\rho$ & Relevance & OOVR & eCPM \\ 
\midrule
Exp.1 & OnlineModel & 59.93\% & 85.41\% & 0.02 & 68.15\% & 0\% & 8676 \\
Exp.2 & DeepNB & 44.16\% & 79.82\% & 0.13 & 46.26\% & 0\% & 12852 \\
Exp.3 & BEQUE & 39.26\% & 62.85\% & 0.08 & 74.37\% & 38.41\% & 7582 \\
Exp.4 & BEQUE+Trie & 42.77\% & 65.36\% & 0.04 & 72.96\% & 0\% & 7225 \\
Exp.5 & BEQUE+WeightedTrie & 53.94\% & 79.20\% & 0.26 & 70.73\% & 0\% & 19275 \\
Exp.6 & raw Qwen2-7B & 12.88\% & 18.37\% & -0.03 & 59.81\% & 68.63\% & 2937 \\
Exp.7 & SFT Qwen2-7B & 38.18\% & 63.71\% & 0.02 & 76.33\% & 34.85\% & 7439 \\
Exp.8 & SFT+Trie & 59.11\% & 86.38\% & -0.01 & 75.31\% & 0\% & 7270 \\
Exp.9 & SFT+DPO+Trie & 38.61\% & 62.19\% & 0.05 & 69.48\% & 0\% & 28391 \\
Exp.10 & \textbf{SFT+WeightedTrie} & 60.37\% & 91.52\% & 0.46 & 74.55\% & 0\% & 59803 \\
Exp.11 & SFT+DPO+WeightedTrie & 50.35\% & 77.73\% & 0.23 & 70.18\% & 0\% & 38301 \\
Exp.12 & \textbf{SFT+WDPO+WeightedTrie(\textit{VALUE})} & \underline{62.67\%} & \underline{91.91\%} & \underline{0.56} & 73.49\% & 0\% & \underline{63775} \\
\bottomrule
\end{tabular}
\end{table*}

In this section, we present the results of offline experiments conducted to evaluate the performance of our proposed models against several baseline models. The comparative performance analysis is summarized in Table \ref{table:performance}. We compared our models with the following baselines: first, the \textbf{Online Model}, which represents the current state-of-the-art multi-channel recall system deployed online; second, the \textbf{DeepNB} model \cite{chen2020rpm}, which employs a vector-based approach focusing solely on bidword value; and finally, the \textbf{BEQUE} model \cite{taobaowww}, which utilizes a LLM fine-tuned with multi-task SFT and PRO alignment\cite{song2024preference}, representing the generative query rewriting approach. The performance metrics used for evaluation include hitrate, eCPM, Spearman's $\rho$, Relevance, and OOVR(Out-Of-Vocabulary Rate. To a certain extent, it is also equivalent to the hallucination rate of LLMs.). These metrics provide a comprehensive assessment of the models' effectiveness in terms of query rewriting, relevance, and OOVR.

\subsubsection{Comprehensive Evaluation of the VALUE Framework}
As evidenced in Table \ref{table:performance}, VALUE (Exp.12) demonstrates superior performance across all evaluation metrics, underscoring the robustness and effectiveness of our proposed method in generating high-value and contextually relevant bidwords for e-commerce query rewriting. DeepNB (Exp.2) leverages click logs for efficiency-oriented rewriting to maximize value, yet its rewrites suffer from compromised relevance to the original search queries. Conversely, the generative approach BEQUE (Exp.3) capitalizes on the semantic comprehension capabilities of LLMs, achieving high rewrite relevance, but exhibits poor value awareness regarding bidwords. In contrast to both methods, VALUE significantly enhances the eCPM and hitrate of the rewritten queries while maintaining high relevance. This indicates that VALUE effectively synthesizes the complex semantic understanding derived from LLMs with value awareness enabled by the Weighted Trie. Notably, BEQUE exhibits a remarkably high Out-of-Vocabulary Word Rate (OOVR) of 38.41\%,  and integrating the Trie-based inference structure into BEQUE (Exp.4) significantly enhances hitrate while ensuring output authenticity. Furthermore, compared to the mainstream generative method "SFT+Trie+DPO" (Exp.9), VALUE achieves improvements across all evaluation metrics. This outcome suggests that our proposed reward-guided inference framework mitigates potential overfitting inherent in the value alignment stage.

\subsubsection{Ablation Study of the Weighted Trie.}
Our proposed Weighted Trie guides LLMs during the generation phase to produce bidwords that optimize both relevance and value, while adapting to real-time changes in their monetary value. Compared to "SFT+Trie" (Exp.8), the "SFT+WeightedTrie" configuration (Exp.10) achieves a 5.1 percentage point increase in hitrate@500 and substantially improves eCPM while concurrently maintaining rewrite relevance. Similarly, when employed in the generation phase alongside other generative rewriting methods (Exp.4 vs Exp.5, Exp.9 vs Exp.11), Weighted Trie significantly enhances key evaluation metrics associated with bidwords value. These results demonstrate that our Weighted Trie serves as an efficient, lightweight, and pluggable module that effectively enhances the value awareness of generative query rewriting methods. Notably, the incorporation of Weighted Trie significantly elevates the Spearman's $\rho$, demonstrating that its value-relevance balancing mechanism during decoding enhances fine-grained value awareness.

\subsubsection{Ablation Study of WDPO.}
Traditional preference alignment techniques employed in BEQUE, such as DPO and Preference Ranking Optimization (PRO), are susceptible to significant overfitting during training, degrading performance in terms of both hitrate and relevance. The overfitting issue inherent to DPO arises from its tendency to overfit specific preference pairs in the training data, leading to compromised generalization capabilities. Consistent with this observation, Table \ref{table:performance} demonstrates marked declines in both hitrate and relevance when integrating DPO with either the "SFT+Trie" (Exp.8 vs Exp.9) or the "SFT+WeightedTrie" (Exp.10 vs Exp.11). In contrast, WDPO enables LLMs to focus learning efforts on high-value bidwords by attenuating the influence of preference pairs exhibiting minimal value differences during training. Consequently, the VALUE framework incorporating WDPO (Exp.12) achieves substantial gains across multiple value-centric metrics while preserving relevance.


\begin{figure}[t]
    \centering
    \includegraphics[width=0.5\textwidth]{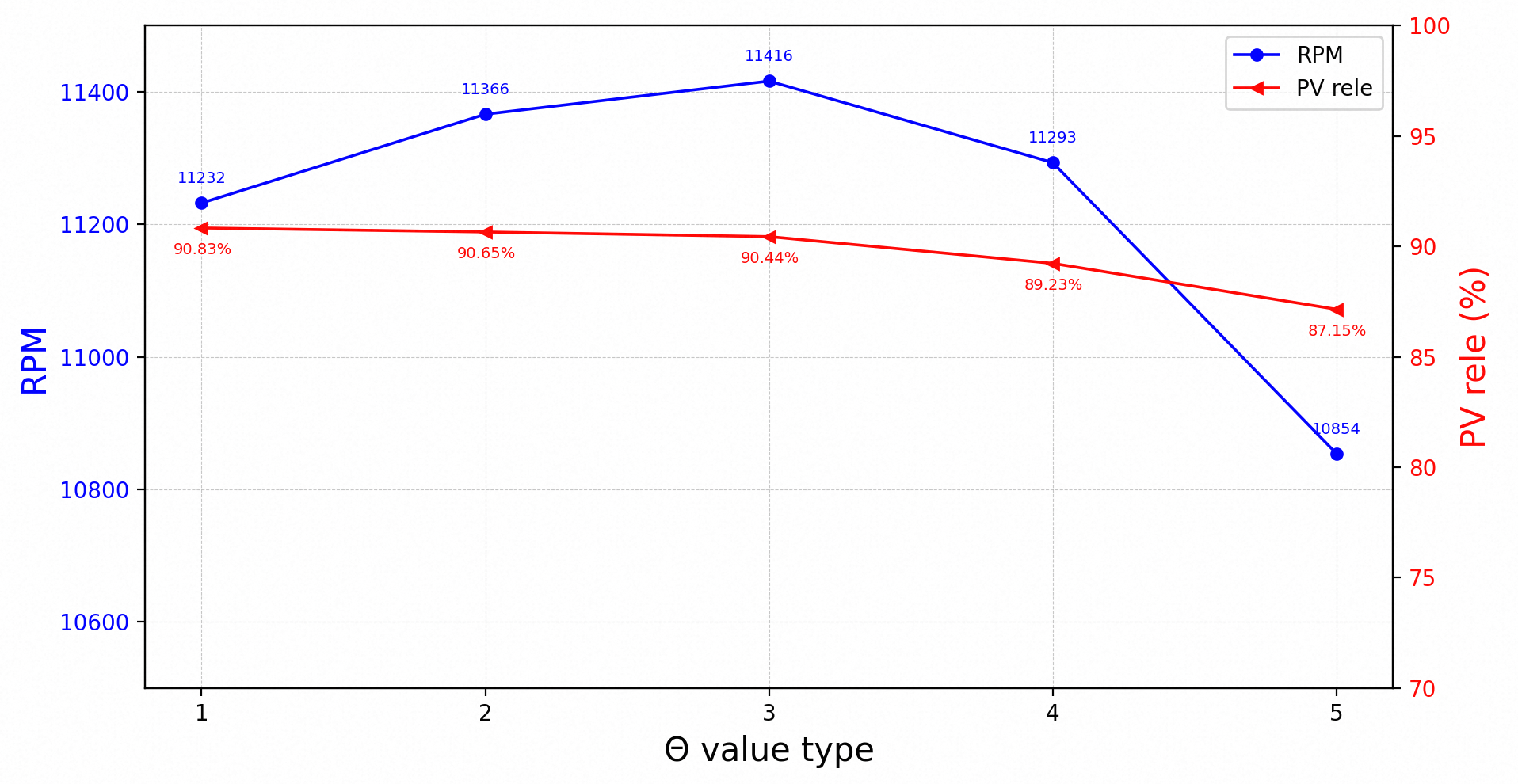}
    \caption{Trade-off between RPM and PV rele through $\Theta$. The specific values of RPM and PV rele are confidential, so the RPM and PV rele here are equivalently scaled.}
    \Description{}
    \label{fig:overview}
\end{figure}

\begin{table}[t]
\centering
\caption{Comparative Performance Analysis of Hyperparameter.}
\label{table:thetaperformance}
\begin{tabular}{lcc}
\toprule
Hyperparameter($\Theta$) & hitrate@500 & Relevance \\ 
\midrule
$\Theta_1$: [0, 0, 0, 0, 0, 0, 0, …] & 86.38\% & 75.31\% \\
$\Theta_2$: [1, 1, 1, 1, 1, 1, 1, …] & 89.65\% & 75.04\% \\
$\Theta_3$: [1, 2, 3, 4, 5, 6, 7, …] & 91.52\% & 74.55\% \\
$\Theta_4$: [1, 2, 4, 8, 16, 32, 64, …] & 88.63\% & 69.85\% \\
$\Theta_5$: [2, 4, 8, 16, 32, 64, 128, …] & 73.71\% & 60.53\% \\
\bottomrule
\end{tabular}
\end{table}

\subsection{Exploration of Hyperparameter Configurations}
In this section, we conduct a thorough investigation into the impact of varying the hyperparameter $\Theta$ on system performance. This parameter, which scales with the depth of the Weighted Trie, modulates the trade-off between relevance and commercial value at different stages of the inference process. Specifically, we evaluate five distinct configurations (see Table \ref{table:thetaperformance}):
\begin{itemize}
    \item \textbf{$\Theta_1$}: Assigns a value weight of 0 throughout, effectively disabling the Weighted Trie and relying solely on the original LLM inference output.
    \item \textbf{$\Theta_2$}: Assigns equal weighting to relevance and value at every depth of the Weighted Trie.
    \item \textbf{$\Theta_3$}: Increases the value weight linearly as a function of Weighted Trie depth.
    \item \textbf{$\Theta_4$}: Increases the value weight exponentially with Weighted Trie depth, significantly prioritizing value in the deeper inference stages.
    \item \textbf{$\Theta_5$}: Further increases the baseline value weights at all depths established by $\Theta_4$.
\end{itemize}

Table \ref{table:thetaperformance} and Figure \ref{fig:overview} present the performance metrics under different $\Theta$ settings across both online and offline evaluations. Key observations reveal that increasing the value weight consistently compromises relevance. Critically, however, excessive value weighting (as in $\Theta_4$ and $\Theta_5$) not only severely degrades relevance but also diminishes RPM and hitrate. Within the sponsored search context, the objective of query rewriting is to maximize the value of rewrites while satisfying a minimum relevance criterion. The $\Theta_3$ configuration addresses this by directing VALUE to prioritize highly relevant bidwords during the initial token generation stages, subsequently shifting focus to high-value bidwords later in the process. This strategy strikes an optimal balance between relevance and value by first ensuring the generation of a more relevant prefix. Our findings underscore that meticulous hyperparameter tuning is paramount for aligning model behavior with specific sponsored search objectives.

\subsection{Scaling Laws in Model Performance}
In our investigation of model performance across varying scales, we conducted experiments using models ranging from 0.5 billion to 65 billion parameters. Our findings consistently demonstrate that larger models exhibit superior performance across all evaluated metrics. Specifically, as model size increases, there is a notable improvement in hitrate, indicating enhanced capability in generating high-value and relevant bidwords. This trend aligns with the theoretical expectations of scaling laws, which suggest that larger models can capture more complex patterns and nuances in data, thereby improving overall effectiveness.

\begin{table}[t]
\centering
\caption{Comparative Performance Analysis of Base Model Size.}
\label{table:basemodelsizeperformance}
\begin{tabular}{lcc}
\toprule
Model Size & hitrate@500 & Relevance \\ 
\midrule
0.5B & 68.32\% & 61.85\% \\
1.5B & 87.49\% & 71.29\% \\
7B & 91.52\% & 74.55\% \\
14B & 92.69\% & 76.68\% \\
65B & 93.98\% & 77.99\% \\
\bottomrule
\end{tabular}
\end{table}

\subsection{Online Experiments}

To further validate the effectiveness of our proposed VALUE framework, we conducted 14-day online A/B tests on our system, focusing on cost and RPM. Results are summarized in Table \ref{table:online} and Table \ref{table:length}.

\begin{table}[t]
\centering
\caption{Online A/B Test Results}
\label{table:online}
\begin{tabular}{lccc}
\toprule
Query Type & Cost & RPM & PV rele \\
\midrule
VALUE  all queries &  +1.53\% & +1.64\% & +0.32pt \\
VALUE  head queries & +0.97\% & +1.03\% & +0.10pt \\
VALUE torso queries & +1.68\% & +1.74\% & +0.33pt \\
VALUE tail queries & +2.35\% & +2.66\%  & +0.75pt \\
\bottomrule
\end{tabular}
\end{table}

The online A/B tests corroborate the offline findings, consistently demonstrating improvements in both cost and RPM across all query types. Our framework exhibits particularly notable enhancements for torso and tail queries, which are typically characterized by greater variability and complexity. This underscores the framework's robust generalization capabilities. As illustrated in Table \ref{table:length}, our model excels in discerning the intent behind very long and complex queries. The observed improvements in RPM and relevance for medium to super long queries underscore the model's proficiency in managing intricate and challenging query structures.

\begin{table}[t]
\centering
\caption{Comparative Performance Analysis of Query Length Layer.}
\label{table:length}
\begin{tabular}{lccccccc}
\toprule
Query Length& Cost & RPM & Relevance \\ 
\midrule
Short Query &+0.41\% & +0.56\% & +0.02pt \\
Medium Query &+0.88\% & +0.98\% & +0.30pt  \\
Long Query &+1.26\% &+1.56\% & +0.79pt  \\
Super Long Query&+1.17\% & +1.55\% & +0.54pt  \\
\bottomrule
\end{tabular}
\end{table}

\subsection{Case Study}  
\begin{figure}[t]
    \centering
    \includegraphics[width=0.5\textwidth]{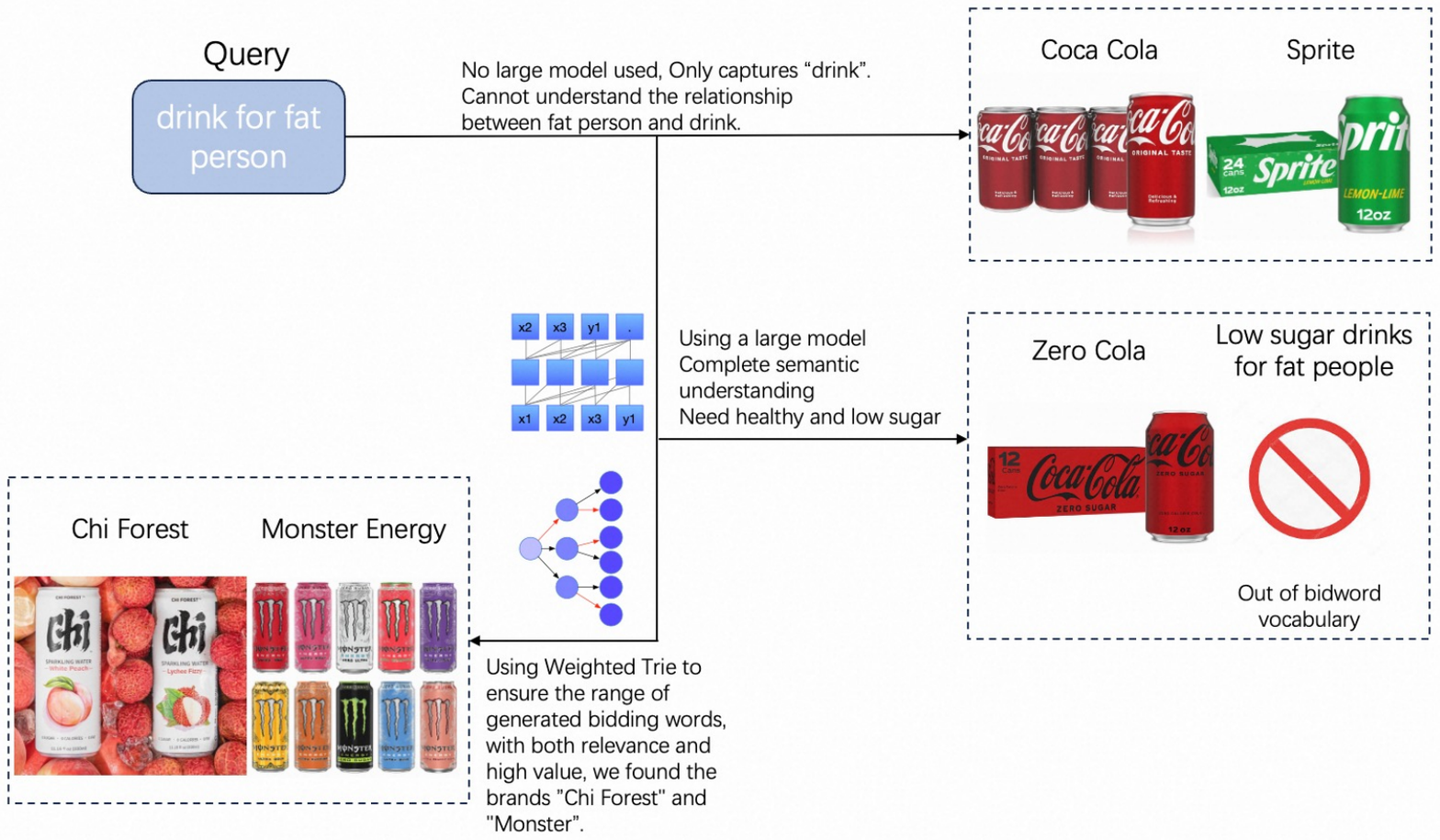}
    \caption{Case Study of VALUE Framework}
    \Description{}
    \label{fig:case}
\end{figure}
In this section, as illustrated in Figure \ref{fig:case}, we examine a specific case involving the user query "drink for fat person". Previously, without the use of LLMs, our system could only match the query at the keyword level, specifically "drink" . With the integration of LLMs, we gained a deeper understanding of the user's intent, identifying a preference for low-sugar, low-calorie beverages. Additionally, by employing a Weighted Trie, we ensured that the advertisements presented were not only closely aligned with the user's intent but also of high commercial value.

\section{Conclusion}
This study introduces \textbf{VALUE}, an innovative framework that addresses the critical challenge of producing rewrites that simultaneously optimize semantic relevance and commercial value, which is a critical limitation pervasive in existing methods. VALUE incorporates real-time value signals into LLM decoding through a novel Weighted Trie and applies WDPO-based alignment that prioritizes bidword pairs with large value gaps during training. These components jointly guide generation toward high-value rewrites without compromising semantic relevance.

Extensive experiments confirm the framework’s effectiveness. Offline tests yield substantial eCPM gains while preserving relevance, and online A/B experiments report significant improvements in cost and Revenue Per Mille. The modular design of VALUE also supports other multi-reward tasks, for example video recommendation in which click-through rate and watch-time completion must be balanced within a fixed generation space and dynamically weighted objectives.



\bibliographystyle{ACM-Reference-Format}
\bibliography{sample-sigconf}

\end{document}